\documentclass[11pt]{article}
\usepackage[pdfmenubar=false, letterpaper=false, pdfpagelabels=true]{hyperref}
\usepackage{epsfig}

\newcommand{\bm}[1]{\mbox{\boldmath$#1$}}
\def\mvec#1{{\bm{#1}}}   

\begin{document}

\title{Fits, and especially linear fits, with errors on both axes,\\
       extra variance of the data points and other complications}
\author{G. D'Agostini
\mbox{} \\
{\small Universit\`a ``La Sapienza'' and INFN, Rome, Italy} \\
{\small (giulio.dagostini@roma1.infn.it, 
\url{http://www.roma1.infn.it/~dagos})}
}
\date{}

\maketitle
\vspace{1.5cm}
\begin{abstract}
The aim of this paper, triggered by some discussions in the astrophysics 
community raised by 
\href{http://arxiv.org/abs/astro-ph/0508529}{astro-ph/0508529}, 
is to introduce the issue of `fits' from a probabilistic perspective
(also known as Bayesian), with special attention to the
construction of model that describes the 
`network of dependences' (a Bayesian network)
that connects experimental observations
to model parameters and upon which the probabilistic inference 
relies. 
The particular case of linear fit
with errors on both axes and extra variance of the data points around 
the straight line (i.e. not accounted by the experimental errors) 
is shown in detail. Some questions related to the
use of linear fit formulas to log-linearized exponential and
power laws are also sketched, as well as the issue of systematic errors.
\end{abstract}

\vspace{1.7cm}
\section*{Preamble}
This paper, based on things already written somewhere
with the addition
of  some details from lectures, 
contains nothing or little especially new. Even the main `result',
summarized in Eq.(\ref{eq:f_mcS}) and that I hope will contribute
to set down the questions raised by 
\href{http://arxiv.org/abs/astro-ph/0508529}{astro-ph/0508529}%
~\cite{astro-ph/0508529}, is just a simple extension of 
Eq.~(8.33) of Ref.~\cite{BR}. Therefore the debated 
question could be dismissed with a paper even shorter than 
\href{http://arxiv.org/abs/astro-ph/0508529}{astro-ph/0508529}.
Nevertheless, I have taken the opportunity to reorganize old
material for the benefit of my students, and 
I post these pages 
hoping they could be of some utility to those
who wish to understand what there is behind formulas. 

\newpage
\section{Introduction}
A common task in data analysis is to `determine',
on the basis of experimental observations, the values of 
the parameters of a model that relates physical quantities. 
This procedure is usually 
associated to names like `fit' and `regression', and to {\it principles},
like  'least squares' or `maximum likelihood' (with variants). 
I prefer, as many others belonging to a still small minority,
to approach the problem from more fundamental probabilistic 
`first principles', that are indeed the fundamental rules 
of probability theory. This approach is also called `Bayesian'
because of the central role played by Bayes' theorem in the process of
learning from data, as we shall see in a while (for 
a critical introduction to the Bayesian approach see 
Ref.~\cite{BR} and references therein).
In practice this means
that we rank in probability hypotheses and 
numerical values about 
which we are not certain.
This is rather intuitive and it is indeed the natural way physicists
reason (see e.g. Ref.~\cite{Fermibayes} and references therein),
though we have been taught a 
peculiar view of probability that does not allow us to make
the reasonings we intuitively do and that
we are going to use here.

In the so called Bayesian approach 
the issue of `fits' takes the name of {\it parametric
inference}, in the sense we are interested in inferring 
the parameters of a model that relates `true' values.
The outcome of the inference is an uncertain 
knowledge of parameters, whose possible values are ranked 
using the language and the tools of probability theory.
As it {\it can only be} 
(see e.g. Ref.~\cite{BR} for extensive discussions),
the resulting inference depends on the inferential model and 
on previous knowledge about the possible values the 
model parameters can take
(though this last dependence is usually rather weak if the inference
is based on a `large' number of observations). It is then important
to state clearly the several assumptions that enter the data analysis.
I hope this paper does it with the due care 
-- and I apologize in advance for some pedantry and repetitions.  
The main message I would like to convey 
is that nowadays it is much more important to build up
the model that describes at best the physics case than 
to obtain simple formulae for the 'best estimates' and their
uncertainty.
This is because, thanks to
the extraordinary progresses of applied mathematics and 
computing power, in most cases 
 the calculation of the integrals that come from a straight application of the
probability theory does not require any longer titanic efforts. 
Building up the correct model is then equivalent, in most cases,
to have solved the problem.

The paper is organized as follows. In Section 
\ref{sec:inference} the inferential approach is introduced from scratch,
only assuming  the multivariate extensions
of the following well known 
formulas\footnote{The meaning of the overall conditioning $I$ 
will be clarified later. Note that, 
in order to simplify the notation, the generic 
symbol $f(\,)$ is used to indicate all probability density functions, 
though they might 
 refer to different variables and have different mathematical
expressions. 
In particular, the order of the arguments is irrelevant, 
in the sense that $f(x,y\,|\,I)$ stands for `joint probability density function
of $x$ and $y$ under condition $I$', 
and therefore it could be also indicated by $f(y,x\,|\,I)$.
For the same reason, the indexes of sums and products and
the extremes of the integrals
are usually omitted, implying they extend
to all possible values of the variables.}
\begin{eqnarray}
f(x,y\,|\,I) &=& f(x\,|\,y,I)\cdot f(y\,|\,I) \label{eq:condprob}\\
f(x\,|\,I)   &=& \int\!f(x,y\,|\,I)\,dy\,.\label{eq:margxy}
\end{eqnarray}
We show how to build the general model, and how this evolves
as soon as the several hypotheses of the model are introduced
(independence, normal error functions, linear dependence 
between true values, vague priors). 
The graphical representation of the model in terms
of the so called `Bayesian networks' is also shown, the utility
of which will become self-evident. 
The case of linear fit with errors on both 
axes is then summarized in Section \ref{sec:lfit_xy}, and the approximate
solution for the non-linear case is sketched in Section \ref{sec:approx}.
The extra variability of the data is modeled in Section
\ref{sec:extra_v}, first in general and then in the 
simple case of the linear fit. The interpretation of the 
inferential result is discussed  in Section \ref{sec:summary_limits},
in which approximated methods to calculate the
 fit summaries (expected values
and variance of the parameters) are shown.
Finally, some comments on the not-trivial issues 
related to the use of linear fit formulas
to infer the parameters of exponential and power laws
are given in Section \ref{sec:linearization}. 
Section \ref{sec:syst} shows how to 
extend the model to include systematic errors, and 
some simple formulas to take into account offset and scale 
systematic errors
in the case of linear fits will be provided.
The paper ends with some conclusions and some comments
about the debate that has triggered it.

\section{Probabilistic parametric inference from a set of 
data points with errors on both axes}\label{sec:inference}
Let us consider a `law' that relates the `true' values of two
quantities, indicated here by
$\mu_x$ and $\mu_y$:
\begin{equation}
\mu_y = \mu_y(\mu_x;\mvec\theta)\,,
\label{eq:muXmuX}
\end{equation}
where $\mvec\theta$ stands for the parameters
of the law, whose number is $M$. 
In the linear case Eq.~(\ref{eq:muXmuX}) reduces to
\begin{eqnarray}
\mu_y &=& m\,\mu_x + c 
\end{eqnarray}
i.e. $\mvec\theta = \{m,c\}$ and $M=2$. 
As it is well understood, because of `errors'
we do not observe directly  $\mu_x$ and $\mu_y$,
but experimental quantities\footnote{These quantities might also be 
summaries of the data. I.e. they are either directly observed numbers,
like readings on scales, or quantities calculated from direct observations,
like averages or other `statistics' based on partial analysis of the data. 
It is implicit that when summaries are used, instead of direct observations,
the analyzer is somewhat relying on the so called 'statistical sufficiency'.} 
$x$ and $y$ that might differ,
on an event by event basis,  
from $\mu_x$ and $\mu_y$. The outcome of 
the `observation' (see footnote 2) 
$x_i$ for a given $\mu_{x_i}$ (analogous reasonings
apply to $y_i$ and $\mu_{y_i}$) 
is modeled by 
an error function $f(x_i\,|\,\mu_{x_i},I)$, that is indeed
a probability density function (pdf) conditioned by 
$\mu_{x_i}$ and the `general state of knowledge' $I$.
The latter stands for all background knowledge  behind the analysis, 
that is what for example makes us to believe 
the relation $\mu_y = \mu_y(\mu_x;\mvec\theta)$, 
the particular mathematical expressions 
for  $f(x_i\,|\,\mu_{x_i},I)$ and $f(y_i\,|\,\mu_{y_i},I)$, 
and so on. Note that 
the shape of the error function might depend on the value of 
$\mu_{x_i}$, 
as it happens if the detector does not respond the same way
to different solicitations. 
A usual {\it assumption} is that 
{\it errors are normally distributed}, i.e.
\begin{eqnarray}
x_i & \sim & {\cal N}(\mu_{x_i}, \sigma_{x_i}) \label{eq:x_norm}\\
y_i & \sim & {\cal N}(\mu_{y_i}, \sigma_{y_i})\,, \label{eq:y_norm}
\end{eqnarray}
where the symbol `$\sim$' stands for `is described by the distribution'
(or `follows the distribution'),
and  where we still leave the possibility that the standard deviations, 
that we consider known, 
might be different in different observations.
Anyway, for sake of generality, we shall make use of 
assumptions (\ref{eq:x_norm}) and (\ref{eq:y_norm}) 
only in next section. 

If we think of $N$ pairs of measurements of $\mu_x$ and $\mu_y$, 
before doing the experiment we are uncertain about $4 N$ quantities
(all $x$'s, all $y$'s, all $\mu_x$'s and all $\mu_y$'s, indicated respectively
as $\mvec x$, $\mvec y$, $\mvec \mu_x$ and $\mvec \mu_y$)
plus the number of parameters, i.e. in total
$4 N + M$, that become $4 N + 2$
in linear fits. [But note that, due to believed deterministic 
relationship (\ref{eq:muXmuX}), the number of independent variables
is in fact $3 N + M$.] Our final goal,
expressed in probabilistic terms, is to get the pdf 
of the parameters given the experimental information
and all background knowledge:
$$ \hspace{5.6cm}\Longrightarrow f(\mvec\theta\,|\,\mvec{x},\mvec{y},I) 
   \ \ \ \ [\,\rightarrow f(m,c\,|\,\mvec{x},\mvec{y},I)
    \ \ \mbox{for linear fits}\,]\,. $$
Probability theory teaches us how to get
the conditional pdf  
$f(\mvec\theta\,|\,\mvec{x},\mvec{y},I)$
if we know  the joint distribution  
$f(\mvec{x},\mvec{y},\mvec{\mu}_x,\mvec{\mu}_y,\mvec\theta\,|\,I)$.
The first step consists in calculating the $2\,N + M$ variable pdf 
(only $N + M$ of which are independent)
that describes the uncertainty of what is not precisely known, given what 
it is (plus all background knowledge). This is achieved 
by a multivariate extension of Eq.~(\ref{eq:condprob}):
\begin{eqnarray}
  f(\mvec{\mu}_x,\mvec{\mu}_y,\mvec{\theta}\,|\,\mvec{x},\mvec{y},I)
&=& \frac{f(\mvec{x},\mvec{y},\mvec{\mu}_x,\mvec{\mu}_y,\mvec{\theta}\,|\,I)}
         {f(\mvec{x},\mvec{y}\,|\,I)} \label{eq:Bayes1}\\
&=& \frac{f(\mvec{x},\mvec{y},\mvec{\mu}_x,\mvec{\mu}_y,\mvec{\theta}\,|\,I)}
         {\int f(\mvec{x},\mvec{y},\mvec{\mu}_x,\mvec{\mu}_y,\mvec{\theta}\,|\,I)  
             \,\,d\mvec{\mu}_x\,d\mvec{\mu}_y\,d\mvec{\theta}}\label{eq:Bayes2} 
\end{eqnarray}
Equations (\ref{eq:Bayes1}) and (\ref{eq:Bayes2}) are two different
ways of writing Bayes' theorem in the case of multiple inference.
Going from (\ref{eq:Bayes1}) to (\ref{eq:Bayes2}) we have `marginalized'
$f(\mvec{x},\mvec{y},\mvec{\mu}_x,\mvec{\mu}_y,\mvec{\theta}\,|\,I)$ over
$\mvec{\mu}_x$, $\mvec{\mu}_y$ and $\mvec{\theta}$, i.e. 
we used an extension of 
Eq.~(\ref{eq:margxy}) to many variables.
[The standard text book version of the Bayes formula differs from 
Eqs.~(\ref{eq:Bayes1}) and (\ref{eq:Bayes2}) because the joint 
pdf's that appear on the r.h.s. of Eqs.~(\ref{eq:Bayes1})-(\ref{eq:Bayes2})
are usually factorized using the so called 
'chain rule', i.e. an extension of Eq.~(\ref{eq:condprob})
to many variables.]

\begin{figure}
\begin{center}
\epsfig{file=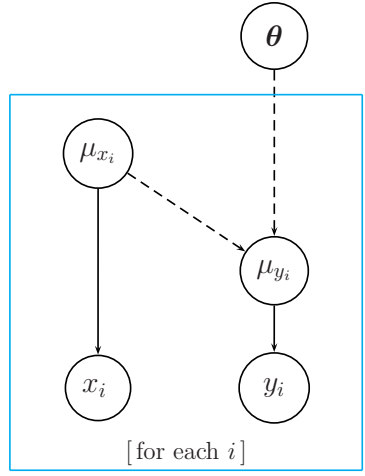,clip=,width=0.3\linewidth}
\end{center}
\caption{{\sl Graphical representation of the model in term of a
Bayesian network (see text).}}
\label{fig:bn1}
\end{figure}

The second step consists in 
marginalizing the $(2\,N + M)$-dimensional pdf over the variables
we are not interested to:
\begin{eqnarray}
f(\mvec{\theta}\,|\,\mvec{x},\mvec{y},I) &=& 
\int\!   f(\mvec{\mu}_x,\mvec{\mu}_y,\mvec{\theta}\,|\,\mvec{x},\mvec{y},I)\,\,
       d\mvec{\mu}_x\,d\mvec{\mu}_y
\label{eq:marginalization}
\end{eqnarray}
Before doing that, we note that the denominator of the r.h.s. 
of Eqs.~(\ref{eq:Bayes1})-(\ref{eq:Bayes2})  is
just a number, once the model and the 
set of observations $\{\mvec{x},\mvec{y}\}$ is defined, and then we can 
 absorb it in the normalization constant. 
Therefore Eq.~(\ref{eq:marginalization}) can be simply rewritten as 
\begin{eqnarray}
f(\mvec{\theta}\,|\,\mvec{x},\mvec{y},I) &\propto& 
\int\!   f(\mvec{x},\mvec{y},\mvec{\mu}_x,\mvec{\mu}_y,\mvec{\theta}\,|\,I)\,\,
       d\mvec{\mu}_x\,d\mvec{\mu}_y\,.
\label{eq:marginalization2}
\end{eqnarray}
We understand then that, essentially, we need to set up 
$f(\mvec{x},\mvec{y},\mvec{\mu}_x,\mvec{\mu}_y,\mvec{\theta}\,|\,I)$
using the pieces of information that come from our 
background knowledge $I$. 
This seems a horrible task, but it becomes feasible tanks to 
the chain rule of probability theory, that allows us to rewrite
$f(\mvec{x},\mvec{y},\mvec{\mu}_x,\mvec{\mu}_y,\mvec{\theta}\,|\,I)$
in the following way:\\ \newpage
\begin{eqnarray}
f(\mvec{x},\mvec{y},\mvec{\mu}_x,\mvec{\mu}_y,\mvec{\theta}\,|\,I) &=& 
  f(\mvec{x}\,|\,\mvec{y},\mvec{\mu}_x,\mvec{\mu}_y,\mvec{\theta},I) \nonumber \\
 &&  \!\!\cdot f(\mvec{y}\,|\,\mvec{\mu}_x,\mvec{\mu}_y,\mvec{\theta},I)  \nonumber \\
  && \!\!  \cdot f(\mvec{\mu}_y\,|\,\mvec{\mu}_x,\mvec{\theta},I) \nonumber \\
 &&\!\! \cdot f(\mvec{\mu}_x\,|\,\mvec{\theta},I) \nonumber \\
  &&\!\! \cdot  f(\mvec{\theta}\,|\,I)
\end{eqnarray}
(Obviously, among the several possible ones, we choose the factorization
that matches our knowledge about of physics case.)
At this point let us make the inventory of the ingredients,
stressing their effective conditions
and making use of independence, when it holds.
\begin{itemize}
\item
Each observation $x_i$ depends directly only on the corresponding true value
$\mu_{x_i}$:
\begin{eqnarray}
 f(\mvec{x}\,|\,\mvec{y},\mvec{\mu}_x,\mvec{\mu}_y,\mvec{\theta},I) 
&=&  f(\mvec{x}\,|\,\mvec{\mu}_x,I) = \prod_i f(x_i\,|\,\mu_{x_i},I) \\
 &&  [\  \Longrightarrow \prod_i {\cal N}(\mu_{x_i},\sigma_{x_i}) \ ].
\end{eqnarray}
(In square brackets is the `routinely' used pdf.)
\item
Each observation $y_i$ depends directly only on the corresponding true value
$\mu_{y_i}$:
\begin{eqnarray}
 f(\mvec{y}\,|\,\mvec{\mu}_x,\mvec{\mu}_y,\mvec{\theta},I) 
   &=&  f(\mvec{y}\,|\,\mvec{\mu}_y,I) = \prod_i f(y_i\,|\,\mu_{y_i},I) \\
 &&  [\  \Longrightarrow \prod_i {\cal N}(\mu_{y_i},\sigma_{y_i}) \ ].
\end{eqnarray}
\item
Each true value $\mu_y$ depends only, and in a deterministic way,
 on the corresponding true value  $\mu_x$
and on the parameters $\mvec{\theta}$. This is
formally equivalent to take an infinitely sharp distribution of 
$\mu_{y_i}$ around $\mu_y(\mu_{x_i};\mvec\theta)$, 
i.e. a Dirac delta function:
\begin{eqnarray}
 f(\mvec{\mu}_y\,|\,\mvec{\mu}_x,\mvec{\theta},I) &=& 
      \prod_i \delta[\,\mu_{y_i}- \mu_{y}(\mu_{x_i},\mvec{\theta})\,] \\
&&  [\  \Longrightarrow \prod_i  \delta( \mu_{y_i} -m\, \mu_{x_i} - c )\ ]
\end{eqnarray}
\item 
Finally, $\mu_{x_i}$ and $\mvec{\theta}$ are usually independent and 
become the {\it priors} of the 
problem,\footnote{Priors need to be specified for the nodes
of a Bayesian network that have no {\it parents} 
(see Fig~\ref{fig:bn1} and footnote~4).
Priors
are {\it logically necessary} ingredients, without 
which probabilistic inference is simply impossible. 
I understand that those who approach 
this kind of reasoning for the first time
might be scared of this `subjective ingredient',
and because of it
they might prefer methods advertised as `objective' 
to which they are used, formally not depending on priors. 
However, if one thinks a bit deeper to the question,
one realizes that behind the slogan of `objectivity' 
there is much arbitrariness,
of which the users are often not aware, and that might lead to 
seriously wrong results in critical problems. Instead, the Bayesian approach
offers the logical tool to properly 
blend prior judgment
and empirical evidence. For further comments see Ref.~\cite{BR},
where it is shown with theoretical arguments and many examples
what is the role of priors,
when they can be `neglected' (never logically! -- 
but almost always in routine data analysis),
and even when they are so crucial that
it is better to refrain from
providing probabilistic conclusions.} 
that one takes `vague' enough, unless physical motivations suggest to do
otherwise. For the $\mu_{x_i}$ we take immediately uniform distributions
over a large domain (a `flat prior').  
Instead, we leave here the expression of $f(\mvec{\theta}\,|\,I)$ undefined,
as a reminder for critical problems (e.g. one of the parameter is positively
defined because of its physical meaning), 
though it can also be taken flat in routine
applications with `many' data points. 
\begin{eqnarray}
    f(\mvec{\mu}_x\,|\,\mvec{\theta},I)\cdot f(\mvec{\theta}\,|\,I) 
     &=&  f(\mvec{\mu}_x\,|\,I)\cdot f(\mvec{\theta}\,|\,I)  \label{eq:f_mux_k}\\
     &=& k_x\,f(\mvec{\theta}\,|\,I) 
\end{eqnarray}
The constant value of $f(\mvec{\mu}_x\,|\,I)$, 
indicated here by $k_x$,   
is then in practice absorbed in the normalization constant. 
\end{itemize}
In conclusion we have
\begin{eqnarray}
f(\mvec{x},\mvec{y},\mvec{\mu}_x,\mvec{\mu}_y,\mvec{\theta}\,|\,I) 
&=& \prod_i f(x_i\,|\,\mu_{x_i},I) \cdot f(y_i\,|\,\mu_{y_i},I) 
  \cdot\, \delta[\,\mu_{y_i}- \mu_{y}(\mu_{x_i},\mvec{\theta})\,]\,
 \cdot f(\mu_{x_i}\,|\,I)\cdot f(\mvec{\theta}\,|\,I)\nonumber\\
&&  \label{eq:model1_norm} \\
&=& \prod_i k_{x_i}\,f(x_i\,|\,\mu_{x_i},I) \cdot f(y_i\,|\,\mu_{y_i},I) 
  \cdot\, \delta[\,\mu_{y_i}- \mu_{y}(\mu_{x_i},\mvec{\theta})\,]\,
 \cdot f(\mvec{\theta}\,|\,I) \label{eq:model1_norm_kxi}\\
&\propto & 
 \prod_i f(x_i\,|\,\mu_{x_i},I) \cdot f(y_i\,|\,\mu_{y_i},I) 
  \cdot\, \delta[\,\mu_{y_i}- \mu_{y}(\mu_{x_i},\mvec{\theta})\,]\,
 \cdot f(\mvec{\theta}\,|\,I)  \,.  \label{eq:model1}
\end{eqnarray}
Figure \ref{fig:bn1}
provides a graphical representation of
the model [or, more precisely, a 
graphical representation of Eq.~(\ref{eq:model1_norm})].
In this diagram the probabilistic connections are indicated 
by solid lines and the deterministic connections by 
dashed lines. 
These kind of networks of probabilistic and deterministic 
relations among uncertain quantities is known as 
`Bayesian network',\footnote{According to 
\href{http://en.wikipedia.org/wiki/Bayesian_network}%
{\it Wikipedia}\,\cite{Wikipedia}, a  
Bayesian network ``{\sl   is a directed graph of nodes representing
 variables and arcs representing dependence relations among the 
variables. If there is an arc from node A to another node B, 
then we say that A is a parent of B. If a node has a known value, 
it is said to be an evidence node. A node can represent any kind 
of variable, be it an observed measurement, a parameter, 
a latent variable, or a hypothesis. 
Nodes are not restricted to representing random variables; 
this is what is "Bayesian" about a Bayesian network.}'' 
[Note: here ``random variable'' stands for a random variable
in the frequentistic acceptation of the term 
(`\`a la von Mises` randomness) and not just as `variable of uncertain
value'.] Bayesian networks represent both a conceptual and a practical
tool to tackle complex inferential problems. They have indeed renewed 
the interest in the field of artificial intelligence, where
they are used in inferential engines, expert systems and decision makers. 
Browsing the web you will find plenty of applications. Here just a few
references: Ref.~\cite{Heckerman} is a well known tutorial;
Ref.~\cite{Jensen} and \cite{Neapolitan} 
and good general books on the subject, 
the first of which is related
to the {\it HUGIN} software, a lite version of it can be 
freely downloaded~\cite{hugin}; for a flash introduction to the
issue, with the possibility of starting playing with Bayesian
network on discrete problems JavaBayes~\cite{jb} is recommended,
for which I have worked also a couple of examples in 
\cite{mybn}; for discrete and continuous variables that can
be modeled with well known pdf, a good starting point is 
{\it BUGS}~\cite{BUGS}, for which I have worked out some examples
concerning uncertainties in measurements~\cite{myBUGS}. 
BUGS stands for {\bf B}ayesian inference {\bf U}sing 
{\bf G}ibbs {\bf S}ampling. This means the relevant integrals
we shall see later are performed by sampling, i.e. 
using Markov chain Monte Carlo (MCMC) methods. 
I do not try to introduce them here, and I suggest to look
elsewhere. Good starting point can be the BUGS web page~\cite{BUGS}
and Ref.~\cite{Andrieu}.
} 
'belief network', 'influence network', 'causal network'
and other names meaning substantially the same 
thing.
From  Eqs.~(\ref{eq:marginalization2}) and (\ref{eq:model1}) we get then 
\begin{eqnarray}
f(\mvec{\theta}\,|\,\mvec{x},\mvec{y},I) &\propto& 
 \left[\int   
  \prod_i k_{x_i}\,f(x_i\,|\,\mu_{x_i},I) \cdot f(y_i\,|\,\mu_{y_i},I) 
         \cdot \delta[\,\mu_{y_i}- \mu_{y}(\mu_{x_i},\mvec{\theta})\,]
        \ d\mvec{\mu}_x\,d\mvec{\mu}_y \right]
     \cdot  f(\mvec{\theta}\,|\,I) \nonumber  \\
&& \label{eq:marginalization3a} \\
 &\propto&  f( \mvec{x},\mvec{y}\,|\,\mvec{\theta},I)  
  \cdot  f(\mvec{\theta}\,|\,I)
 =  {\cal L}(\mvec{\theta}\,;\,  \mvec{x},\mvec{y})
        \cdot  f(\mvec{\theta}\,|\,I)
\label{eq:marginalization3}
\end{eqnarray}
where we have factorized the unnormalized `final' pdf into  the 
`likelihood'\footnote{Traditionally the name `likelihood' is
given to the probability of the data given the parameters, i.e. 
$f( \mvec{x},\mvec{y}\,|\,\mvec{\theta},I)$, seen as a 
mathematical function of the parameters. Therefore the notation
${\cal L}(\mvec{\theta}\,;\, \mvec{x},\mvec{y})$
[not to be confused with $f(\mvec{\theta}\,|\, \mvec{x},\mvec{y})$!]. 
$f( \mvec{x},\mvec{y}\,|\,\mvec{\theta},I)$ can be obtained marginalizing 
$f(\mvec{x},\mvec{y},\mvec{\mu}_x,\mvec{\mu}_y\,|\,\mvec{\theta},I)$, i.e.
$f( \mvec{x},\mvec{y}\,|\,\mvec{\theta},I) = 
\int f(\mvec{x},\mvec{y},\mvec{\mu}_x,\mvec{\mu}_y\,|\,\mvec{\theta},I)
\, d\mvec{\mu}_x
d\mvec{\mu}_x$, 
where 
$f(\mvec{x},\mvec{y},\mvec{\mu}_x,\mvec{\mu}_y\,|\,\mvec{\theta},I)=
f(\mvec{x},\mvec{y},\mvec{\mu}_x,\mvec{\mu}_y,\mvec{\theta}\,|\,I) 
  / f(\mvec{\theta}\,|\,I)$ is obtained from Eq.~(\ref{eq:model1_norm}). 
 It follows:
\begin{eqnarray*}
f(\mvec{\mvec{x},\mvec{y},\mu}_x,\mvec{\mu}_y\,|\,I) 
&=& \prod_i f(x_i\,|\,\mu_{x_i},I) \cdot f(y_i\,|\,\mu_{y_i},I) 
  \cdot\, \delta[\,\mu_{y_i}- \mu_{y}(\mu_{x_i},\mvec{\theta})\,]\,
 \cdot f(\mu_{x_i}\,|\,I) 
\end{eqnarray*}
and 
\begin{eqnarray*}
f( \mvec{x},\mvec{y}\,|\,\mvec{\theta},I) &=& 
\int \prod_i f(x_i\,|\,\mu_{x_i},I) \cdot f(y_i\,|\,\mu_{y_i},I) 
  \cdot\, \delta[\,\mu_{y_i}- \mu_{y}(\mu_{x_i},\mvec{\theta})\,]\,
 \cdot f(\mu_{x_i}\,|\,I) \ d\mu_{x_i}d\mu_{y_i}.
\end{eqnarray*}
} 
${\cal L}(\mvec{\theta}\,;\, \mvec{x},\mvec{y})$
(the content of the large square bracket)  and  the `prior'  
$f(\mvec{\theta}\,|\,I)$. 

We see than that, a part from the prior,  the result is essentially 
given by the product of $N$ terms, each 
of which depending on the individual pair of measurements:
\begin{eqnarray}
f(\mvec{\theta}\,|\,\mvec{x},\mvec{y},I) &\propto&  
                \left[\prod_i {\cal L}_i(\mvec{\theta}\,;\,x_i,y_i,I)\right]
                \cdot f(\mvec{\theta}\,|\,I)\,, \label{eq:lik_prior} 
\end{eqnarray} 
where
\begin{eqnarray}
{\cal L}_i(\mvec{\theta}\,;\,x_i,y_i) = f(x_i,y_i\,|\,\mvec{\theta},I)  &=& 
k_{x_i}\, \int  f(x_i\,|\,\mu_{x_i},I) \cdot f(y_i\,|\,\mu_{y_i},I) 
 \cdot \delta[\,\mu_{y_i}- \mu_{y}(\mu_{x_i},\mvec{\theta})\,] \,\,
 d{\mu_{x_i}}d{\mu_{y_i}} \nonumber \\
 && \label{eq:lik_i_0} \\
   &=& 
k_{x_i}\, \int  f(x_i\,|\,\mu_{x_i},I) \cdot f(y_i\,|\,\mu_{y}(\mu_{x_i},\mvec{\theta}),I) 
\,\, d{\mu_{x_i}} 
\end{eqnarray}
and the constant factor $k_{x_i}$, irrelevant in the Bayes formula, 
is a reminder of the priors about  $\mu_{x_i}$ (see footnote 5). 

\section{Linear fit with normal errors on both axes}\label{sec:lfit_xy}
To apply the general formulas of the previous section
 we only need to make explicit 
$\mu_{y_i}(\mu_{x_i},\mvec{\theta})$
 and the error functions, and finally 
integrate over $\mu_{x_i}$. 
In the case of linear fit with normal errors 
the individual contributions 
to the likelihoods become 
\begin{eqnarray}
{\cal L}_i(m,c\,;\,x_i,y_i)  &=& k_{x_i}
\int  \frac{1}{\sqrt{2\pi}\, \sigma_{x_i}}\, 
\exp{ \left[ -\frac{(x_i-\mu_{x_i})^2}
                   {2\,\sigma_{x_i}^2} 
           \right] }
  \cdot  \frac{1}{\sqrt{2\pi}\, \sigma_{y_i}}\, 
\exp{ \left[ -\frac{(y_i-m\,\mu_{x_i}-c)^2}
                   {2\,\sigma_{y_i}^2} 
      \right]
} \,\, d{\mu_{x_i}}\, \nonumber \\ && \label{eq:lik_i_1}\\
 &=&  k_{x_i} \, \frac{1}{\sqrt{2\pi}\, \sqrt{\sigma_{y_i}^2+m^2\,\sigma_{x_i}^2}}\,
 \exp{ \left[ -\frac{(y_i-m\,x_i-c)^2}
                   {2\, (\sigma_{y_i}^2+m^2\,\sigma_{x_i}^2) } 
      \right]}\,,
\end{eqnarray}
that, inserted into Eq.~(\ref{eq:lik_prior}), finally give 
\begin{eqnarray}
f(m,c\,|\,\mvec{x},\mvec{y},I) &\propto& \prod_i  
 \frac{1}{\sqrt{\sigma_{y_i}^2+m^2\,\sigma_{x_i}^2}}\,
 \exp{ \left[ -\frac{(y_i-m\,x_i-c)^2}
                   {2\, (\sigma_{y_i}^2+m^2\,\sigma_{x_i}^2) } 
      \right]}\,  f(m,c\,|\,I)\,. 
\label{eq:f_mc}
\end{eqnarray}
The effect of the error of the $x$-values is to have an effective
standard error on the  $y$-values that is the quadratic combination
of $\sigma_y$ and $\sigma_x$, the latter `propagated' 
to the other coordinate via the slope $m$
(this result can be justified heuristically by dimensional analysis).

\section{Approximated solution for non-linear fits with normal errors}
\label{sec:approx}
Linearity implies that the arguments of the exponential 
of the integrand in Eq.~(\ref{eq:lik_i_1}) contains only 
first and second powers of $\mu_{x_i }$, and then the integrals
has a closed solution. Though this is not true in general,
the linear case teaches us how to get an approximated 
solution of the problem. We can take first order expansions
of $\mu_{y}(\mu_{x},\mvec{\theta})$ around each $x_i$
\begin{eqnarray}
\mu_y(\mu_{x_i};\mvec{\theta}) &\approx &
\mu_y(x_i;\mvec{\theta}) \, + \, 
\mu_y^{\,\prime}(x_i;\mvec\theta)\cdot
(\mu_{x_i}-x_i)\,.
\end{eqnarray}
The difference $y_i-m\,\mu_{x_i}-c$ in Eq.~(\ref{eq:lik_i_1}),
that was indeed equal to
 $y_i-\mu_y(\mu_{x_i};\mvec\theta)$ in the general case,
using the linear approximation becomes
$$ y_i - \mu_y(x_i;\mvec{\theta}) - 
\mu_y^{\,\prime}(x_i;\mvec\theta)\cdot (\mu_{x_i}-x_i)
= y_i - \mu_y^{\,\prime}(x_i;\mvec\theta)\cdot \mu_{x_i} - 
        [\,\mu_y(x_i;\mvec{\theta}) -  
         \mu_y^{\,\prime}(x_i;\mvec\theta)\cdot x_i\,]
\,, $$
i.e. 
we have the following replacements in Eqs.~(\ref{eq:lik_i_1})-(\ref{eq:f_mc}):
\begin{eqnarray}
m &\rightarrow& \mu_y^{\,\prime}(x_i;\mvec\theta) \\
c  &\rightarrow& \mu_y(x_i;\mvec{\theta}) -  \mu_y^{\,\prime}(x_i;\mvec\theta)\cdot x_i\,.
\end{eqnarray}
The approximated equivalent of 
Eq.~(\ref{eq:f_mc}) is then 
\begin{eqnarray}
f(\mvec\theta\,|\,\mvec{x},\mvec{y},I) &\propto\approx& \prod_i  
 \frac{1}{\sqrt{\sigma_{y_i}^2+{\mu_y^{\,\prime}}^2
 (x_i;\mvec\theta)\cdot\sigma_{x_i}^2}}\,
 \exp{ \left[ -\frac{[\,y_i-\mu_y(x_i;\mvec\theta)\,]^{\,2}}
                   {2\, [\sigma_{y_i}^2+{\mu_y^{\,\prime}}^2(x_i;\mvec\theta)\cdot\sigma_{x_i}^2] } 
      \right]}\,  f(\theta\,|\,I)\,,\ \ \ \ 
\label{eq:f_gen_approx}
\end{eqnarray}
where the unusual symbol `$\propto\approx$' stands for 
`approximately proportional to'. 

\section{Extra variability of the data}\label{sec:extra_v}
As clearly stated, the previous results assume that the only sources of
deviation  of the measurements from the value of the physical quantities 
are  normal errors,  with known standard deviations $\sigma_{x_i}$ and
 $\sigma_{y_i}$ . 
Sometimes, as it is the case
of the data points reported in Ref.~\cite{astro-ph/0508483}, 
this is not the case. This means that $y$ depends also 
on other, `hidden' variables, and what we observe is the overall
effects integrated over all the variability of the variables that
we do not `see'. In lack of more detailed information, the simplest modification
to the model described above is to add an extra Gaussian
 `noise' on one of the coordinates. 
For tradition and simplicity this extra noise is added 
to the $y$ variable. 
The effect on the above result can be easily understood. Let us call
$\sigma_v$ the r.m.s. of 
this extra noise that acts normally and independently in each  
$y$ point. As it is well known, the sum of Gaussian distributions
is still Gaussian with an expected value and variance respectively 
sum of the individual expected values and variances.  
Therefore, the effect in the individual likelihoods (\ref{eq:lik_i_1}) 
is to replace  $\sigma^2_{y_i}$ by $\sigma^2_{y_i}+\sigma^2_v$. 
But we now have an extra parameter in the model, and Eq.~(\ref{eq:f_mc}) 
becomes
\begin{eqnarray}
f(m,c,\sigma_v\,|\,\mvec{x},\mvec{y},I) &\propto& \prod_i  
 \frac{1}{\sqrt{\sigma^2_v +  \sigma_{y_i}^2+m^2\,\sigma_{x_i}^2}}\,
 \exp{ \left[ -\frac{(y_i-m\,x_i-c)^2}
                   {2\, (\sigma^2_v + \sigma_{y_i}^2+m^2\,\sigma_{x_i}^2) } 
      \right]}\,  f(m,c,\sigma_v\,|\,I)\,.  \nonumber \\
\label{eq:f_mcS}
\end{eqnarray}
More rigorously, this formula can be obtained from a variation
of reasoning followed in the previous section.
\begin{itemize}
\item 
$\mu_y$ depends on $\mu_x$ and on the set of hidden
variables  $\mvec{v}$:
\begin{eqnarray}
\mu_y &=& \mu_y^{(v)}(\mu_x,\mvec{\theta}, \mvec{v}) \\
      &=& z(\mu_x,\mvec{\theta}) + g(\mu_x,\mvec{v})\, 
\end{eqnarray}
where the overall dependence $\mu_y^{(v)}(\,\,)$ 
has been split in two  functions: $z(\mu_x,\mvec{\theta})$,
 only depending on $\mu_x$ and the model parameters,
corresponding to the ideal case; 
$g(\mu_x,\mvec{v})$ describing the difference from the 
ideal case.
\item
Calling $z$ the fictitious variable, deterministically dependent 
on $\mu_x$, for a given $\mu_{x_i}$ we have the following model
\begin{eqnarray}
z_i = z(\mu_{x_i}, \mvec{\theta}) \,:&& f(z_i\,|\,\mu_{x_i}, \mvec{\theta},I)
=  \delta[\,z_i - z(\mu_{x_i},\mvec{\theta})\,] \\
\mu_{y_i} \,: && f(\mu_{y_i}\,|\,z_i,I)\,     
\end{eqnarray}
where $f(\mu_{y_i}\,|\,z_i,I)$ describes our uncertainty about 
$\mu_{y_i}$ due to the unknown values of all other hidden variables. 
\item
We need now to specify $f(\mu_{y_i}\,|\,z_i,I)$. As usual, 
in lack of better knowledge, we take a Gaussian distribution 
of unknown parameter $\sigma_v$, 
with awareness that this is just a convenient, approximated 
way to quantify our uncertainty. 

At this point a summary of all ingredients
of the model in the specific case of linear model is in order: 
\begin{eqnarray}
  y_i & \sim & {\cal N}(\mu_{y_i}, \sigma_{y_i}) \label{eq:mf1}\\
  x_i & \sim & {\cal N}(\mu_{x_i}, \sigma_{x_i}) \label{eq:mf2}\\
  z_i & \leftarrow & m\, \mu_{x_i} + c 
               \hspace{10.0mm} [\,\Rightarrow\ \delta(z_i- m\, \mu_{x_i} + c)\,]
   \label{eq:mf3}    \\
  \mu_{y_i} &  \sim & {\cal N}(z_i, \sigma_{v}) \label{eq:mf4} \\
  \mu_{x_i}  & \sim & {\cal U}(-\infty,+\infty)  
  \hspace{5.0mm} [\,\Rightarrow\  k_{x_i}\,] \label{eq:mf5} \\
  m,c,\sigma_v &\Rightarrow & 
\mbox{see later}  \hspace{12.4mm} [\,\Rightarrow \mbox{'uniform'}\,], \label{eq:mf6}
\end{eqnarray}
where ${\cal U}(-\infty,+\infty)$ stands for a uniform distribution 
over a very large interval, and the symbol `$\leftarrow$' 
has been used to deterministically assign a value, as done in 
BUGS\,\cite{BUGS} (see later).
\item
We have now the extra parameter $\sigma_v$ that we include in $\mvec\theta$,
so that $M$ increases by 1. 
The new model in represented in Fig.~\ref{fig:bn2}, 
\begin{figure}
\begin{center}
\epsfig{file=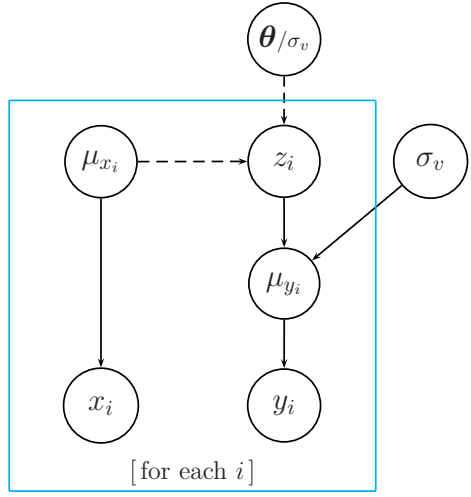,clip=,width=0.39\linewidth}
\end{center}
\caption{{\sl Minimal modification of Fig.~\ref{fig:bn1} 
to model the extra variability not described by the error functions. 
Note that $\mvec\theta$ stands for all model parameters to be inferred, 
including $\sigma_v$. Instead, $\mvec\theta/\sigma_v$ stands for 
all parameters apart from $\sigma_v$.}}
\label{fig:bn2}
\end{figure}
in which 
we have indicated by $\mvec\theta/\sigma_v$ all parameters apart
from $\sigma_v$.
\item
The variables of the model are now $5 N+M$, and Eq.~(\ref{eq:model1}) 
becomes
\begin{eqnarray}
f(\mvec{x},\mvec{y},\mvec{\mu}_x,\mvec{\mu}_y,\mvec{z},\mvec{\theta}\,|\,I) &\propto & 
 \prod_i f(x_i\,|\,\mu_{x_i},I) \cdot f(y_i\,|\,\mu_{\mu_i},I) \nonumber \\
 &&\ \ \  \cdot f(\mu_{y_i}\,|\,z_i,I)
  \cdot\, \delta[\,z_i - z(\mu_{x_i},\mvec{\theta})\,]\,
 \cdot f(\mvec{\theta}\,|\,I)  \,.  \label{eq:model2}
\end{eqnarray}
\item 
Consequently, Eq.~(\ref{eq:marginalization2}) becomes 
\begin{eqnarray}
f(\mvec{\theta}\,|\,\mvec{x},\mvec{y},I) &\propto& 
\int\!   f(\mvec{x},\mvec{y},\mvec{\mu}_x,\mvec{\mu}_y,\mvec{z},
         \mvec{\theta}\,|\,I)\,\,
         d\mvec{\mu}_x\,d\mvec{\mu}_y\,d\mvec{z}\,.
\label{eq:marginalization4}
\end{eqnarray}
\item
Inserting the model functions (\ref{eq:mf1})-(\ref{eq:mf6}) in 
Eq.~(\ref{eq:model2}), after the marginalization 
(\ref{eq:marginalization4}) and the factorization of the result
into likelihood as prior 
[as previously done in (\ref{eq:marginalization3})], we get 
the analogues of Eqs.~(\ref{eq:lik_i_0})-(\ref{eq:lik_i_1}):
\begin{eqnarray}
\frac{{\cal L}_i(\mvec{\theta}\,;\,x_i,y_i)}{k_{x_i}} &=& 
\int\!  f(x_i\,|\,\mu_{x_i},I) \cdot f(y_i\,|\,\mu_{y_i},I) 
 \cdot f( \mu_{y_i}\,|\,z_i,\mvec{\theta},I) 
 \cdot \delta[\,z_i- z(\mu_{x_i},\mvec{\theta})\,] \,\,
 d{\mu_{x_i}}d{\mu_{y_i}}d{z_i} \nonumber \\
 & & \label{eq:lik_i_2} \\
   &=& 
\int\!  f(x_i\,|\,\mu_{x_i},I) \cdot 
      f(y_i\,|\,\mu_{y_i},I) \cdot
f(\mu_{y_i}\,|\, z(\mu_{x_i},\mvec{\theta}),\mvec{\theta},I) 
\,\, d\mu_{x_i}d\mu_{y_i} \\
 &=& \int  \frac{1}{\sqrt{2\pi}\, \sigma_{x_i}}\, 
\exp{ \left[ -\frac{(x_i-\mu_{x_i})^2}
                   {2\,\sigma_{x_i}^2} 
           \right] } \cdot
 \frac{1}{\sqrt{2\pi}\, \sigma_{y_i}}\, 
\exp{ \left[ -\frac{(y_i-\mu_{y_i})^2}
                   {2\,\sigma_{y_i}^2} 
           \right] } 
\nonumber \\
 &&  \hspace{3.0mm}\cdot  \frac{1}{\sqrt{2\pi}\, \sigma_{v}}\, 
\exp{ \left[ -\frac{(\mu_{y_i}-m\,\mu_{x_i}-c)^2}
                   {2\,\sigma_{v}^2} 
      \right]
} \,\, d\mu_{x_i} d\mu_{y_i}\, \label{eq:lik_i_V} \\
 &=& \int  \frac{1}{\sqrt{2\pi}\, \sigma_{x_i}}\, 
\exp{ \left[ -\frac{(x_i-\mu_{x_i})^2}
                   {2\,\sigma_{x_i}^2} 
           \right] } \nonumber \\
 &&  \hspace{3.0mm} \cdot  \frac{1}{\sqrt{2\pi}\, \sqrt{\sigma_{v}^2+\sigma_{y_i}^2}}\, 
\exp{ \left[ -\frac{(\mu_{y_i}-m\,\mu_{x_i}-c)^2}
                   {2\,(\sigma_{v}^2+\sigma_{y_i}^2)} 
      \right]
} \,\, d\mu_{x_i} \, \label{eq:lik_i_V_I1} \\
 &=&  \frac{1}{\sqrt{2\pi}\, \sqrt{\sigma_{v}^2+\sigma_{y_i}^2
                                   +m^2\sigma_{x_i}^2 } }\, 
\exp{ \left[ -\frac{(\mu_{y_i}-m\,x_i-c)^2}
                   {2\,(\sigma_{v}^2+\sigma_{y_i}^2+m^2\sigma_{x_i}^2)} 
      \right]
} \label{eq:lik_i_V_I2} 
\end{eqnarray}
\item
Inserting in Eq.~(\ref{eq:lik_prior}) the expression of
 ${\cal L}_i(\mvec{\theta}\,;\,x_i,y_i)$ coming from 
Eq.~(\ref{eq:lik_i_V_I2})
 we get finally Eq.~(\ref{eq:f_mcS}).
\end{itemize}

\section{Computational issues: normalization, fit summaries, priors and 
approximations}
\label{sec:summary_limits}
At this point it is important to understand that in Bayesian 
approach the full result of the inference is given by final 
distribution, that in our case is -- we rewrite it here:
\begin{eqnarray}
f(m,c,\sigma_v\,|\,\mvec{x},\mvec{y},I) &=& k\, \prod_i  
 \frac{1}{\sqrt{\sigma^2_v +  \sigma_{y_i}^2+m^2\,\sigma_{x_i}^2}}\,
 \exp{ \left[ -\frac{(y_i-m\,x_i-c)^2}
                   {2\, (\sigma^2_v + \sigma_{y_i}^2+m^2\,\sigma_{x_i}^2) } 
      \right]}\,  f(m,c,\sigma_v\,|\,I)\,,  \nonumber \\
\label{eq:f_mcS_k}
\end{eqnarray}
where $k$ is  `simply' a normalization factor. 
(This factor is usually the most difficult thing to calculate 
and it is often obtained approximately by numerical methods. 
But this is, in principle, just a technical issue.) Once we have got $k$
we have a full knowledge about $f(m,c,\sigma_v\,|\,\mvec{x},\mvec{y},I)$
and therefore about our uncertainty concerning the model parameters, 
the distribution of each of which can be obtained by marginalization:
\begin{eqnarray}
f(m\,|\,\mvec{x},\mvec{y},I) &=& 
  \int\! f(m,c,\sigma_v\,|\,\mvec{x},\mvec{y},I)\,\,dc\,d\sigma_v \\
f(c\,|\,\mvec{x},\mvec{y},I) &=& 
  \int\! f(m,c,\sigma_v\,|\,\mvec{x},\mvec{y},I)\,\,dm\,d\sigma_v \\
f(\sigma_v\,|\,\mvec{x},\mvec{y},I) &=& 
  \int\! f(m,c,\sigma_v\,|\,\mvec{x},\mvec{y},I)\,\,dm\,dc\,.
\end{eqnarray}
Similarly the joint distribution of $m$ and $c$ can be obtained as
\begin{eqnarray}
f(m,c\,|\,\mvec{x},\mvec{y},I) &=& 
  \int\! f(m,c,\sigma_v\,|\,\mvec{x},\mvec{y},I)\,d\sigma_v\,,
\end{eqnarray}
from which we can easily see that we recover Eq.~(\ref{eq:f_mc}) 
in the case we think the extra variability discussed in the previous
section is absent. This limit case corresponds  to a prior of $\sigma_v$
sharply peaked around zero, i.e. $f(\sigma_v\,|\,I) = \delta(\sigma_v)$. 

Other interesting limit cases are the following.
\begin{itemize}
\item 
Errors only on the $y$ axis and no extra variability. \\
Making the limit of Eq.~(\ref{eq:f_mc}) for $\sigma_{x_i}\rightarrow 0$
and neglecting irrelevant factors we get
\begin{eqnarray}
f(m,c\,|\,\mvec{x},\mvec{y},I) &\propto& \prod_i  
 \exp{ \left[ -\frac{(y_i-m\,x_i-c)^2}
                   {2\, \sigma_{y_i}^2} 
      \right]}\,  f(m,c\,|\,I)
\label{eq:f_mc_sx0} \\
 &\propto& \exp{ \left[-\frac{1}{2}\sum_i \frac{(y_i-m\,x_i-c)^2}
                   {\sigma_{y_i}^2}   \right]}\,  f(m,c\,|\,I) \,.
\label{eq:f_mc_sx0_sum}
\end{eqnarray}
This is the best known and best understood case.
\item 
Errors only on the $y$ axis and extra variability. \\
Making the limit of Eq.~(\ref{eq:f_mcS_k}) for $\sigma_{x_i}\rightarrow 0$
\begin{eqnarray}
f(m,c,\sigma_v\,|\,\mvec{x},\mvec{y},I) &\propto& \prod_i  
 \frac{1}{\sqrt{\sigma^2_v +  \sigma_{y_i}^2}}\,
 \exp{ \left[ -\frac{(y_i-m\,x_i-c)^2}
                   {2\, (\sigma^2_v + \sigma_{y_i}^2) } 
      \right]}\,  f(m,c,\sigma_v\,|\,I)\,. 
\label{eq:f_mcS_k_sx0}
\end{eqnarray}
\item 
Scattering of data point around the hypothesized straight line 
only due to `extra variability'. 
\begin{eqnarray}
f(m,c,\sigma_v\,|\,\mvec{x},\mvec{y},I) &\propto & \sigma_v^{-N}  \prod_i  
 \exp{ \left[ -\frac{(y_i-m\,x_i-c)^2}
                   {2\,\sigma^2_v } 
      \right]}\,  f(m,c,\sigma_v\,|\,I)
\label{eq:f_mcS_k_sx0_sy0} \\
 &\propto & \sigma_v^{-N} 
 \exp{ \left[
  -\frac{1}{2\,\sigma_v^2} \,\sum_i  (y_i-m\,x_i-c)^2
 \right]}\,  f(m,c,\sigma_v\,|\,I)
\,.
\end{eqnarray}
This case corresponds to the joint determination of  $m$, $c$ and $\sigma_v$
made by the method of the `residuals', that can be considered
a kind of approximated solution of Eq.~(\ref{eq:f_mcS_k_sx0_sy0}),
achieved by iteration. 
[Indeed, if there are `enough' data points the `best estimates'  
achieved by the residual method are very close to the expected 
values of $m$, $c$ and $\sigma_v$ 
evaluated from $f(m,c,\sigma_v\,|\,\mvec{x},\mvec{y},I)$ if we assumed
a flat prior distribution for the parameters.]
\end{itemize}
Although, as it has been pointed out above,
 the full result of the inference is provided by the final pdf,
often we do not need such a detailed description of our uncertainty,
and we are only interested to provide some `summaries'. The most
interesting ones are the expected values, standard deviations and 
correlation coefficients, i.e. $E(m)$, $E(c)$, $E(\sigma_v)$, 
 $\sigma(m)$, $\sigma(c)$, $\sigma(\sigma_v)$, $\rho(m,c)$, 
$\rho(m,\sigma_v)$ and $\rho(c,\sigma_v)$. They are evaluated from 
$f(m,c,\sigma_v)$ using their definitions, that are assumed 
to be known
[hereon we often omit the conditions on which the pdf
depends, and we write $f(m,c,\sigma_v)$ instead of 
$f(m,c,\sigma_v\,|\,\mvec{x},\mvec{y},I)$, and so on].
Obviously, these are not the only possible summaries. One 
might report in addition the mode or the median of each variable, one-dimensional
or multi-dimensional probability regions [i.e. regions in the space
of the parameters that are believed to contain the true value 
of the parameter(s) with a well defined probability level], 
and so on. It all depends on how standard or unusual the shape
of $f(m,c,\sigma_v)$  is. I just would like to stress that the most important
summaries are expected value, standard deviation and 
correlation coefficients, because these are the quantities that mostly 
matter in subsequent evaluations of uncertainty. Giving only 
`most probable' values and probability intervals might bias the 
results of further analyzes~\cite{asymmetric}.

The prior  $f(m,c,\sigma_v\,|\,I)$ has been left on purpose open
 in the above
formulas, although we have already anticipated that usually a flat
prior about all parameters gives the correct result in most 'healthy'
cases, characterized by a sufficient number of data points. 
I cannot go here through an extensive discussion about the
issue of the priors, often criticized as the weak point of the Bayesian
approach and that are in reality one of its points of force. I refer
to more extensive discussions available elsewhere (see e.g. \cite{BR}
and references therein), giving here only a couple of advices. 
A flat prior is in most times a good starting point (unless
one uses some packages, like BUGS~\cite{BUGS}, that does not like
flat prior in the range $-\infty$ to $+\infty$; in this case 
one can mimic it with a very broad distribution, like a 
Gaussian with very large $\sigma$). 
If the result of the inference `does not offend your physics
sensitivity', it means that, essentially, flat priors have done a good
job and it is not worth fooling around with more sophisticated ones.
In the specific case we are looking closer, that of 
Eq.~(\ref{eq:f_mcS_k}),
the most critical quantity to watch is obviously $\sigma_v$, because 
it is positively defined. If, starting from a flat prior 
(also allowing negative values),
the data constrain the value of $\sigma_v$ 
in a (positive) region far from zero,
and -- in practice consequently -- 
its marginal distribution is approximatively
Gaussian, it means the flat prior was a reasonable choice.
Otherwise, the next-to-simple modeling of $\sigma_v$ is via
the step function $\theta(\sigma_v)$. A more technical choice 
would be a gamma distribution, with suitable parameters
to `easily' accommodate all envisaged values of $\sigma_v$.

The easiest case, that happens very often if one has `many' data points
(where `many' might be already as few as some dozens), 
is that $f(m,c,\sigma_v)$ obtained starting from flat priors
is approximately a multi-variate Gaussian distribution, 
i.e. each marginal is approximately Gaussian. 
In this case the expected value of each variable is 
close to its mode, that, since the prior was a constant, 
corresponds to the value for which the likelihood
${\cal L}(m,c,\sigma_v\,;\, \mvec{x},\mvec{y})$ gets its maximum. 
Therefore the parameter estimates derived by the 
maximum likelihood principle are very good approximations
of the expected values of the parameters calculated directly
from $f(m,c,\sigma_v)$. In a certain sense the maximum likelihood
principle best estimates are recovered as a special case that holds 
under particular conditions (many data points and vague priors).
If either condition fails, the result 
the formulas derived from such a principle 
might be incorrect. 
This is the reason I dislike unneeded principles of this 
kind, once we have a more general framework, of which the 
methods obtained by  `principles' are just special cases under well
defined conditions. 

The simple case in which $f(m,c,\sigma_v)$ is approximately 
multi-variate Gaussian allows also to approximately 
evaluate the covariance matrix of the fit parameters from 
the Hessian of its logarithm.\footnote{I would like to point out
that I added the formulas that follow just for the benefit of the inventory.
Personally, in such low dimensional problems I find it 
easier to perform numerical integrations than to evaluate, 
obviously with the help of some software, derivatives, 
find minima and invert matrices, or to use the 
`$\Delta\,\chi^2=1$' or `$\Delta\,$minus-log-likelihood = $1/2$'
rules. Moreover, I think that the lazy use of
computer programs solely based on some approximations 
produces the bad habit of taking acritically their results, 
even when they make no sense\cite{asymmetric}. Nevertheless, 
with some reluctance and after these warnings, I give 
here the formulas that follows, and that the reader might know
as derived from other ways, hoping he/she understands better
how they can be framed in a more general scheme, and therefore when 
it is possible to use them.}
This is due to a well known 
property of the multi-variate Gaussian and it is not strictly 
related to flat priors. 
In fact it can easily proved that if the generic $f(\mvec\theta)$ 
is a multivariate Gaussian, then 
\begin{eqnarray}
(V^{-1})_{ij}(\mvec\theta) &=& \left.\frac{\partial^2 \varphi}
                                    {\partial\theta_i\,\partial\theta_j}
                          \right|_{\mvec\theta=\mvec\theta_m}
\label{eq:hessian}
\end{eqnarray}
where 
\begin{eqnarray}
\varphi(\mvec\theta) &=& -\log f(\mvec{\theta})\,,
\end{eqnarray}
$V_{ij}(\mvec\theta)$ is the covariance matrix of the parameters
and $\mvec\theta_m$ is the value for which 
$f(\mvec\theta)$
gets its maximum and then $\varphi(\mvec\theta)$ its minimum. 

An interesting feature of this approximated procedure is that, 
since it is based on the logarithm of the pdf, normalization factors
are irrelevant. In particular, if the priors are flat, the 
relevant summaries of the inference
can be obtained from the logarithm of the likelihood, stripped
of all irrelevant factors (that become additive constants
in the logarithm and vanish in the derivatives). 
Let us write down, for some cases of interest, 
the minus-log-likelihoods, stripped of constant terms
and indicated by $L$, i.e. 
$\varphi(\mvec\theta\, ;\, \mvec{x},\mvec{y} ) = 
L(\mvec\theta\, ;\, \mvec{x},\mvec{y} ) + \mbox{\it const}$.
\begin{itemize}
\item 
Simplest case: linear fit with only known errors on the $y$ axis 
[from Eq.~(\ref{eq:f_mc_sx0})]: 
\begin{eqnarray}
L(m,c\,;\, \mvec{x},\mvec{y}) &=& \frac{1}{2}\sum_i\frac{(y_i-m\,x_i-c)^2}
                      {\sigma_{y_i}^2}  =  \frac{1}{2}\, 
\chi^2(m,c\,;\, \mvec{x},\mvec{y})\,,\label{eq:chi2}
\end{eqnarray}
where we recognize the famous chi-squared. 
Applying Eq.~(\ref{eq:hessian}) we get then the covariance matrix
of the fit parameters as 
\begin{eqnarray}
(V^{-1})_{m,c} &=& \left.\frac{1}{2}
                    \frac{\partial^2 \chi^2(m,c\,;\, \mvec{x},\mvec{y})}
                         {\partial m\,\partial c}
                          \right|_{\begin{array}{l} m=m_m \\ c=c_m\end{array}}
\label{eq:V_chi2}
\end{eqnarray}
(See Ref.~\cite{BR} for the fully developed example yielding 
analytic formulas
for the expected values and covariance matrix of the $m$ and $c$.)
Note that the often used 
(but also often misused!\,\cite{asymmetric}) 
`$\Delta\,\chi^2=1$ rule' to calculate the covariance matrix
of the parameters comes from the same Gaussian approximation
of the final pdf and prior insensitivity.
[And, because of the factor $1/2$ between Eqs.~(\ref{eq:hessian}) and 
(\ref{eq:V_chi2}), there is an equivalent 
`$\Delta\,$minus-log-likelihood = $1/2$' rule, applicable under the same 
conditions].
\item 
Errors also on the $y$ axis:
\begin{eqnarray}
L(m,c\,;\, \mvec{x},\mvec{y}) &=& 
\frac{1}{2}\sum_i \log {(\sigma_{y_i}^2+m^2\,\sigma_{x_i}^2)} + 
\frac{1}{2}\sum_i\frac{(y_i-m\,x_i-c)^2}
                      {\sigma_{y_i}^2+m^2\,\sigma_{x_i}^2}\,. \label{eq:eq:L_f_mc_xy}
\end{eqnarray}
In this case expected values and covariance matrix cannot be 
obtained directly in closed form. Nevertheless, 
one can use iteratively the formulas for $\sigma_{x_i}=0$
in which the estimate of $m$ is used to evaluate the terms 
$\sigma_{y_i}^2+m^2\,\sigma_{x_i}^2$ 
(having the meaning of effective $y$-error)
in the likelihood
of the next iteration.
 Instead it is wrong 
to simply replace the denominator of the $\chi^2$ 
of Eq.~(\ref{eq:chi2}) with 
$\sigma_{y_i}^2+m^2\,\sigma_{x_i}^2$, because this approximation does not
take into account the first term of the r.h.s. of Eq.~(\ref{eq:eq:L_f_mc_xy})
and the slope $m$ will be underestimated (as a consequence, the intercept $c$ 
will be over- or under-estimated, depending on the sign of the correlation 
coefficient between $m$ and $c$, 
a sign that depends on the sign of the barycenter of the $x$ points.)
\item
Dispersion on the $y$ axis only due to $\sigma_v$ 
[from Eq.~(\ref{eq:f_mcS_k_sx0_sy0})]:
\begin{eqnarray}
L(m,c,\sigma_v\,;\, \mvec{x},\mvec{y}) &=& 
N\,\log\sigma_v + \frac{1}{2\,\sigma_{v}^2}\sum_i\,(y_i-m\,x_i-c)^2\,. 
\end{eqnarray}
\item
The most complete case seen here [from Eq.~(\ref{eq:f_mcS_k})]:
\begin{eqnarray}
L(m,c,\sigma_v\,;\, \mvec{x},\mvec{y}) &=& \frac{1}{2}\sum_i 
                   \log {(\sigma^2_v +  \sigma_{y_i}^2+m^2\,\sigma_{x_i}^2)}
                + \frac{1}{2}\sum_i \frac{(y_i-m\,x_i-c)^2}
                   {\sigma^2_v + \sigma_{y_i}^2+m^2\,\sigma_{x_i}^2 }\,. 
\label{eq:L_f_mcS_k}
\end{eqnarray}
\item
As the previous item, but for the general $\mu_y(\,)$ 
[from Eq.~(\ref{eq:f_gen_approx})]:
\begin{eqnarray}
L(\mvec\theta,\sigma_v\,;\, \mvec{x},\mvec{y}) 
\!\!\!&\approx&\!\!\! \frac{1}{2}\sum_i 
                   \log {[\sigma^2_v +  \sigma_{y_i}^2
            +{\mu_y^{\,\prime}}^2(x_i; \mvec\theta)\cdot\sigma_{x_i}^2]}
            + \frac{1}{2}\sum_i \frac{[\,y_i-\mu_y(x_i;\mvec\theta)\,]^{\,2}}
                   {\sigma^2_v + \sigma_{y_i}^2
                   +{\mu_y^{\,\prime}}^2
               (x_i; \mvec\theta)\cdot\sigma_{x_i}^2 }\,. 
\nonumber \\ &&\label{eq:L_f_NL_S_k}
\end{eqnarray}
\end{itemize}

\section{From power law to linear fit}\label{sec:linearization}
Linear fits are not only used to infer the parameters of
a linear model, but also of other models that are linearized
via a suitable transformation of the variables. The best known 
cases are the exponential law, linearized taking the log of the ordinate,
and the power low, linearized taking the log of both coordinates. 
Linearizion is particularly important to provide a visual 
evidence in support of the claimed model. However, quantitative 
inference based on the transformed variable is not so obvious,
if high accuracy in the determination of the model parameters
is desired. Let us make some comments on the power law, in 
which both variables are log-transformed and therefore more general.

We start hypothesizing a model 
\begin{eqnarray}
   B & = & \kappa \,A^\gamma\,,
\label{eq:power}
\end{eqnarray}
that is linearized as 
\begin{eqnarray}
   \log B & = & \gamma\, \log A + \log \kappa\,.
\label{eq:powerlin}
\end{eqnarray}
We identify then $\log B$ with $\mu_y$ of the linear case, 
$\log A$ with $\mu_x$, $\gamma$ with $m$ and $\log \kappa$ with $c$. 
But this identification does not allows us yet to use 
{\it tout court} the formulas derived above, because 
each of them depends on a well defined model. 
Let us see where are the possible problems. 
\begin{itemize}
\item
In the simplest model $a_i$ is normally distributed around 
$A_i$ and $b_i$ around $B_i$
(we indicate by  $\mvec{a}$ and $\mvec{b}$ the set of observations
in the original variables).
But, in general, $x_i\equiv \log a_i$
and  $y_i\equiv \log b_i$ are not normally distributed 
around  $\mu_{x_i}\equiv \log A$ and  $\mu_{y_i}\equiv \log B$, respectively. 
They are only when the measurements are very precise, i.e. 
$\sigma_{a_i}/ a_i \ll 1$ and $\sigma_{b_i}/b_i \ll 1$. This the case
in which standard `error propagation', based on the well known 
formulas base on linearization, holds.
\item
If the precision is not very high, i.e. $\sigma_{a_i}/ a_i$
and  $\sigma_{b_i}/b_i$ are not very small, 
non-linear effects in the transformations could be important
(see e.g. Ref.~\cite{asymmetric}). 
\item
When some of $\sigma_{a_i}/ a_i$ and  $\sigma_{b_i}/b_i$ approach
unity it becomes important to consider 
the error functions and the priors about $A$ and $B$
with the due care. For example,  
very often the quantities $A$ and $B$ are defined 
positive -- and if we take their logarithms, they have to be positive. 
This requires the model to be correctly set up in order 
to prevent negative values of $A$ and $B$.
\end{itemize}
Further considerations would require a good knowledge 
of the the experimental apparatus and of the physics
under study. Therefore I refrain from indicating a toy model,
that could be used acritically in serious applications. 
Instead I encourage to draw a graphical representation
of the model, as done in Figs. \ref{fig:bn1} and  \ref{fig:bn2}
and to make the inventory of the ingredients.
Sometimes the representation in terms of Bayesian network
is almost equivalent to solve the problem, thanks 
also to the methods developed in the past decades to calculate the 
relevant integrals, using e.g. Markov Chain Monte Carlo 
(MCMC), see e.g. Ref.~\cite{Andrieu} and references therein. 
In case of simple models one can even use free available 
software, like BUGS~\cite{BUGS}.

\section{Systematic errors}\label{sec:syst}
Let us now consider the effect of systematic errors, 
i.e. errors that acts the same way on all observations
of the sample, for example an uncertain offset in 
the instrument scale, or an uncertain scale factor. 
I do not want to give a complete treatment of the subjects,
but focus only on how our systematic effects 
modify our graphical model,
and give some practical rules for the simple case of linear fits. 
(For an introduction about systematic errors and their consistent
treatment within the Bayesian approach see Ref.~\cite{BR}.)

\begin{figure}[t!]
\begin{center}
\epsfig{file=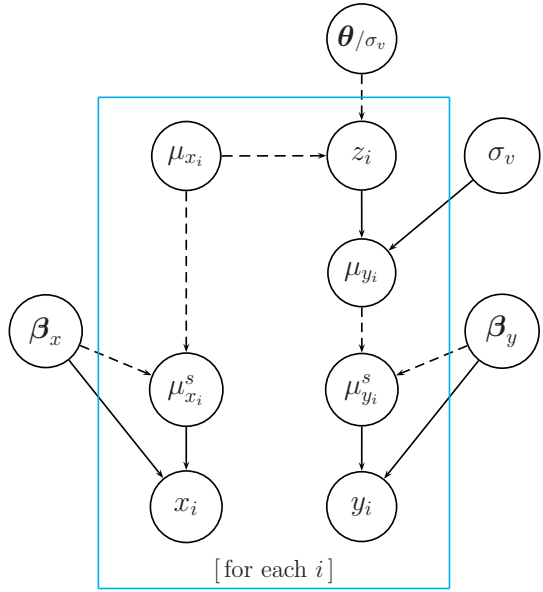,clip=,width=0.45\linewidth}
\end{center}
\caption{{\sl Graphical model of Fig.~\ref{fig:bn2} 
with the addition
of systematic errors on both axes.}}
\label{fig:bn3}
\end{figure}

For each coordinate we can introduce the fictitious quantities
$\mu_x^S$ and $\mu_y^S$ that take into account the 
modification of $\mu_x$ and $\mu_y$ due to the systematic effect.
For example, if the systematic effects only acts as an {\it offset}, 
i.e. we are uncertain about the 
`true' {\it zero} of the instruments, $\zeta_x$ and $\zeta_y$, 
we have 
\begin{eqnarray}
\mu_{x_i}^S &=& \mu_{x_i} + \zeta_x \\
\mu_{y_i}^S &=& \mu_{y_i} + \zeta_y\,,
\end{eqnarray}
where the true value of $\zeta_x$  are $\zeta_y$ 
unknown (otherwise there would be no systematic errors). 
We only know that their expected value is zero (otherwise
we need to apply a calibration constant to the measurements)
and we quantify our uncertainty with pdf's. For example, 
we could model them with Gaussian distributions:
\begin{eqnarray}
\zeta_x & \sim & {\cal N}(0, \sigma_{\zeta_x}) \\
\zeta_y & \sim & {\cal N}(0, \sigma_{\zeta_y})\,.
\end{eqnarray}
Anyway, for sake of generality, we leave the systematic 
effects in the most general form, dependent on the uncertain 
quantities  $\mvec\beta_x$ and $\mvec\beta_y$ [to be clear: 
in the case of solely offset systematics we have
$\mvec\beta_x=\{\zeta_x\}$
$\mvec\beta_y=\{\zeta_y\}$]. 
The values of $\mu_{x_i}^S$ 
and $\mu_{y_i}^S$ are modeled as follow
\begin{eqnarray}
\mu_{x_i}^S \,: &&\ \   \mu_{x_i}^S \leftarrow \mu_{x}^S(\mu_{x_i};\mvec\beta_x) \\
\mu_{y_i}^S  \,: &&\ \   \mu_{y_i}^S \leftarrow \mu_{y}^S(\mu_{y_i};\mvec\beta_y)\\
\mvec\beta_x \, : &&\ \  \mvec\beta_x \sim f(\mvec\beta_x\,|\,I) \\
\mvec\beta_y  \,: &&\ \  \mvec\beta_y \sim f(\mvec\beta_y\,|\,I)\,.
\end{eqnarray}
Figure \ref{fig:bn3} 
\begin{figure}[t!]
\begin{center}
\epsfig{file=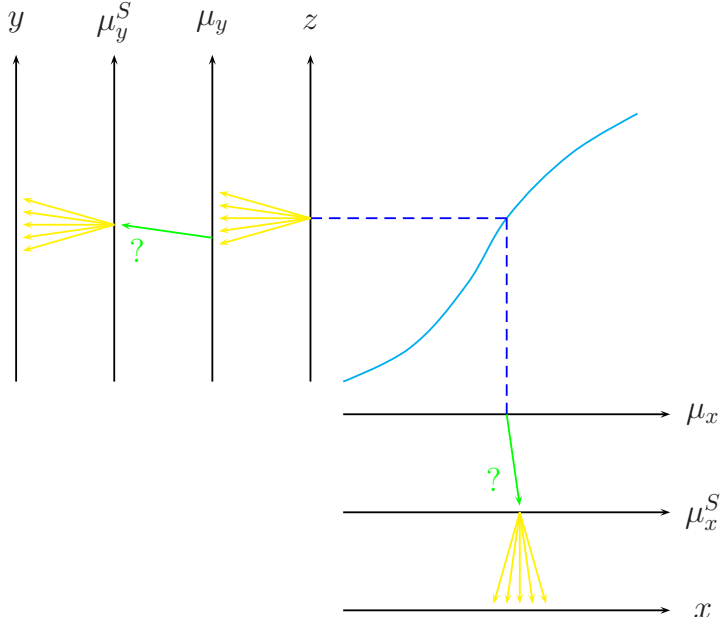,clip=,width=0.6\linewidth}
\end{center}
\caption{{\sl A different visual representation
of the probabilistic model of Fig.~\ref{fig:bn3}.}}
\label{fig:plot}
\end{figure}
shows the graphical model containing the 
new ingredients. 
The links $\mvec\beta_x \rightarrow x_i$
and $\mvec\beta_y \rightarrow y_i$ are to remember that 
systematics could also effect the error functions. 
An alternative visual picture of the 
probabilistic model is shown in Fig.~\ref{fig:plot}. 
Note the different symbols to indicate the different uncertain
processes: the divergent arrows (in yellow, if you are reading 
an electronic version of the paper) indicate that, given a value
of the `parent' variable, the `child' variable fluctuates on an event-by-event
basis; the green single arrow with the question mark indicate that,
given a value of the `parent', the child will always take a fixed value, 
though we do not know which one. 

Obviously, the practical implementation of complicate systematic
effects in complicate fits can be quite challenging, but at least
the Bayesian network provides an overall picture of the model.
The simplest case is that of linear fit where only offset 
and scale uncertainty are present, with uncertainty modeled
by a Gaussian distribution.
This means that the 
$\mvec\beta$'s and their uncertainty are as follows
($\eta$ is the scale factor of uncertain value):
\begin{eqnarray}
\mvec\beta_x = \{\zeta_x,\eta_x\} \hspace{3.0mm}&&
  \hspace{3.0mm} \mvec\beta_y = \{\zeta_y,\eta_y\} \\
\zeta_x \sim {\cal N}(0,\sigma_{\zeta_x})  \hspace{3.0mm} && 
 \hspace{3.0mm}\zeta_y \sim {\cal N}(0,\sigma_{\zeta_y})  \\
\eta_x \sim {\cal N}(1,\sigma_{\eta_x})  \hspace{3.0mm} && 
 \hspace{3.0mm}\eta_y \sim {\cal N}(1,\sigma_{\eta_y})
\end{eqnarray}
In this case we can get 
an hint of how the uncertainty about $m$ and $c$ change without
doing the full calculation
following an heuristic approach, valid when 
$f(m,c)$ is approximately multivariate Gaussian and 
the details of which can be found
in Ref.~\cite{nota1094}. We obtain the following results, 
in which $\left.\sigma(m)\right|_{\zeta_x}$ indicates the contribution 
to the uncertainty about the slope $m$ due to uncertainty
about $\zeta_x$,  $\left.\sigma(m)\right|_{\eta_x}$
that due to the scale factor $\eta_x$, and so 
on\footnote{In Ref.~\cite{nota1094} $\zeta_x$ is indicated by 
$z_x$, $\eta_x$ by $f_x$, and so on.}:\\  
\begin{eqnarray}
\left.\sigma(m)\right|_{\zeta_x} &=& 0 \\
\left.\sigma(m)\right|_{\zeta_y} &=& 0 \\
\left.\sigma(c)\right|_{\zeta_x} &=& |m|\,\sigma_{\zeta_x} \\
\left.\sigma(c)\right|_{\zeta_y} &=& \sigma_{\zeta_y} \\
&& \nonumber \\
\left.\sigma(m)\right|_{\eta_x} &=& |m|\,\sigma_{\eta_x} \\
\left.\sigma(m)\right|_{\eta_y} &=& |m|\,\sigma_{\eta_y} \\
\left.\sigma(c)\right|_{\eta_x} &=& 0 \\
\left.\sigma(c)\right|_{\eta_y} &=& |c|\,\sigma_{\eta_y} \,.
\end{eqnarray}
All contributions are then added quadratically to the 
so called `statistical' ones. 

\section{Conclusions}
The issue of fits has been approached from probability 
first principles, i.e. using throughout the rules
of probability theory, without external {\it ad hoc}
ingredients. It has been that the main task
consists in building up the inferential model,
that means in fact to properly factorize the joint probability
density function of all variables of the problem.
We have seen that this factorization, based on the so called
chain rule of probability theory, has a very convenient
graphical representation, that takes the name of 
Bayesian (or belief/causal/influence) network. 
Modeling the problem in terms of such networks
not only helps to understand the problem better, but, 
thanks the huge amount of mathematical 
developments relates to them, it becomes the only
way to get a (numerical) solution when problems get complicated.

We have also seen how to recover well known formulas, 
obtained starting from other approaches, under well
defined conditions, thus indicating that other methods
can be seen as approximations of the most general one, 
and that are therefore applicable if the conditions of
validity hold.

The linear case with errors on both axis and extra variance 
of the data has been shown with quite some detail,
giving un-normalized formulas for the pdf. In particular, 
going to the pretext to write this paper, we can see that 
Eq.~(43) of Ref.~\cite{astro-ph/9912368} is not reproduced. 
In fact, if I understand it correctly, that equation should have the
same meaning of Eq.~(\ref{eq:f_mcS_k}) of this paper. 
However, Eq.~(43) of Ref.~\cite{astro-ph/9912368} contains
an extra factor $\sqrt{1+m^2}$ (using the notation of this
paper), that it is a bit odd, for several reasons
(besides the fact that I do not get it -- but this could be
judged a technical argument by the hurry reader).
The first reason is just dimensionality:
$m\,x$ is 
homogeneous with 
$y$  and for this reason 
$m\,\sigma_x$ can be combined (quadratically) to  $\sigma_y$, 
but $m^2$ cannot be added {\it tout court} to 1.
The second is that if there was such a factor in  Eq.~(\ref{eq:f_mcS_k}),
then one cannot reproduce Eqs.~(\ref{eq:f_mc_sx0}), 
(\ref{eq:f_mcS_k_sx0}) and (\ref{eq:f_mcS_k_sx0_sy0}), 
that one can be obtained in simpler ways (and that give rise to the
likelihoods shown in Section \ref{sec:summary_limits},
some of them rather well known).
Note that the addition of a term $\sqrt{1+m^2}$ in 
Eq.~(\ref{eq:f_mcS_k}) has the net effect of overestimating $m$,
an effect that is consistent with the claim by \cite{astro-ph/0508529}
of a slope larger than that obtained by  
\cite{astro-ph/0508483}.\footnote{As a rule of thumb, since the extra
variance of the data of \cite{astro-ph/0508483} is rather important, 
the slope has to be very close to that obtained neglecting
all $\sigma_{x_i}$ and  $\sigma_{y_i}$  and making a very
simple least square regression.}


\end{document}